\documentclass{article}
\usepackage{spconf,amsmath,graphicx}

\usepackage{enumitem}
\setlist{nosep, leftmargin=14pt}

\usepackage{mwe} 

\title{Is fitting error a reliable metric for assessing deformable motion correction in quantitative MRI?}

\name{Fanwen Wang\textsuperscript{1,3}, Ke Wen\textsuperscript{2,3}, Yaqing Luo\textsuperscript{2,3}, Yinzhe Wu\textsuperscript{1,3}, Jiahao Huang\textsuperscript{1,3}, Dudley J. Pennell\textsuperscript{2,3}}

\nameone{Pedro F. Ferreira\textsuperscript{2,3}, Andrew D. Scott\textsuperscript{2,3}, Sonia Nielles-Vallespin\textsuperscript{2,3}, Guang Yang\textsuperscript{1,3}.}

\address{$^{1}$Bioengineering Department and {Imperial-X}, Imperial College London, UK.\\
$^{2}$National Heart and Lung Institute, Imperial College London, UK.\\
$^{3}$Cardiovascular Magnetic Resonance Unit, Royal Brompton Hospital, \\Guy’s and St Thomas’ NHS Foundation Trust, UK.}

\begin{document}

\maketitle

%
%
%
%
%
%

%

%
\begin{abstract}
Quantitative MR (qMR) can provide numerical values representing the physical and chemical properties of the tissues. To collect a series of frames under varying settings, retrospective motion correction is essential to align the corresponding anatomical points or features. Under the assumption that the misalignment makes the discrepancy between the corresponding features larger, fitting error is a commonly used evaluation metric for motion correction in qMR. This study evaluates the reliability of the fitting error metric in cardiac diffusion tensor imaging (cDTI) after deformable registration. We found that while fitting error correlates with the negative eigenvalues, the negative Jacobian Determinant increases with broken cardiomyocytes, indicated by helix angle gradient line profiles. Since fitting error measures the distance between moved points and their re-rendered counterparts, the fitting parameter itself may be adjusted due to poor registration. Therefore, fitting error in deformable registration itself is a necessary but not sufficient metric and should be combined with other metrics. 
\end{abstract}
\section{Introduction}
\label{sec:intro}

\begin{figure}[t]
\centering
\includegraphics[width=\columnwidth]{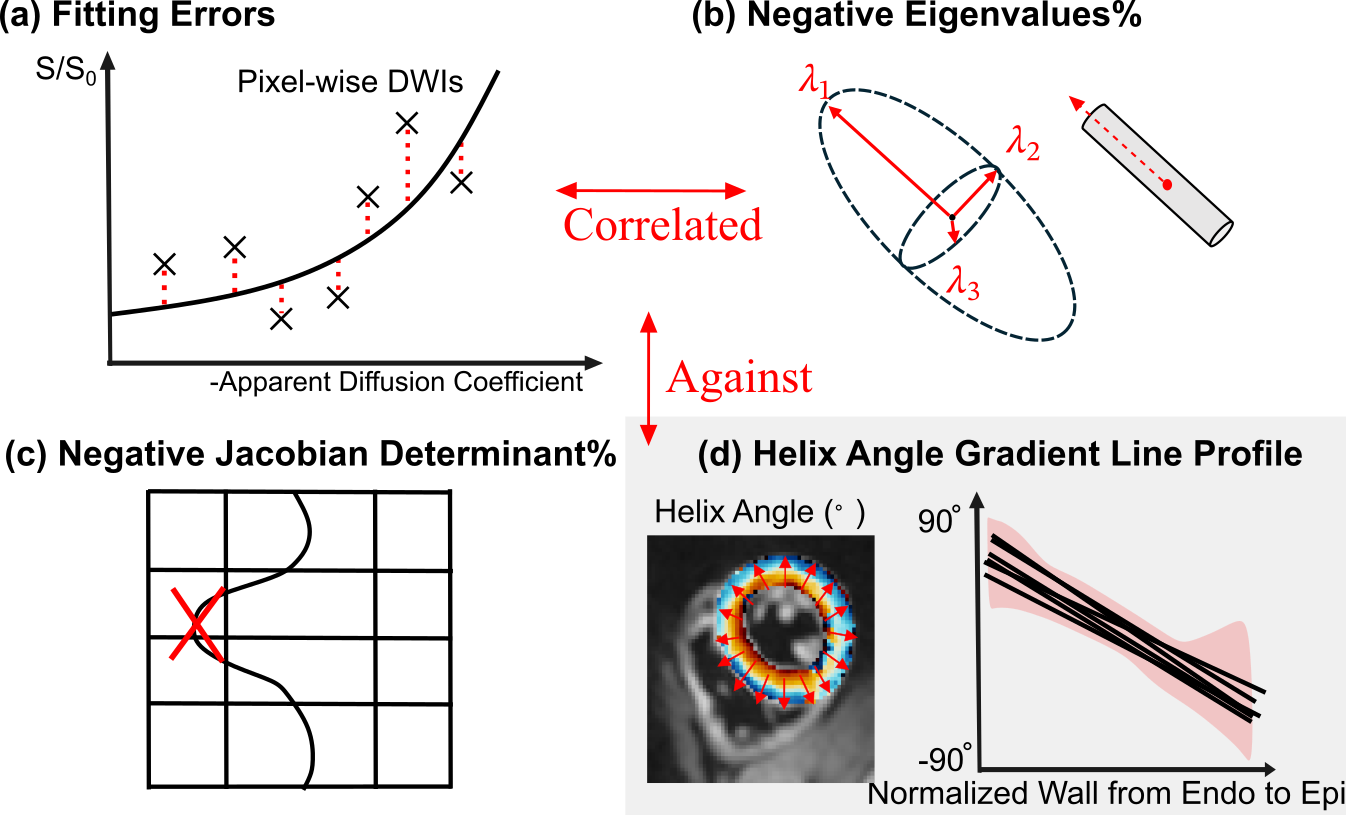}
\caption{The evaluation metrics we used on cDTI.}
\label{EvaluationMetric}
\end{figure}

Quantitative magnetic resonance (qMR) imaging, including T1, T2 mapping and diffusion can provide functional information about the tissues. The core assumption underlying qMR is that the anatomy being scanned remains stationary and unchanged, allowing for consistent measurements across different scanning parameters or sessions. However, cardiac and respiratory motion occurs during different scans with multi-contrast acquisitions. Image registration, with the application of retrospective motion correction, aligns two datasets in one coordinate system. Deformable registration has been used in cardiac mapping \cite{tao2018robust,li2022motion}, kidney \cite{seif2016diffusion}, cardiac diffusion tensor imaging (cDTI) \cite{wang2024groupwise} and abdominal diffusion-weighted imaging \cite{guyader2015influence}. 

Since the mapping between the datasets is usually unknown, the performance of registration relies on the landmark alignment or contour overlaps. The plausibility of the displacement field can also be a surrogate metric. However, these general metrics that work for most of the anatomical images may fail in qMR. The delineation of the contours or annotations of landmarks is labour-intensive and challenging due to the different contrasts and fuzzy boundaries in a series of qMR images. Hence, down-stream fitting process is widely used as an evaluation metric under the assumption that misalignment of the tissues may trigger larger errors \cite{tao2018robust,li2022motion, seif2016diffusion}. However, the reliability of this metric remains unsure.

\begin{table}[ht]
    \centering
    \caption{Evaluation metrics for different registration methods. Models with numbers are deformable registrations with varying regularization weights. \textit{Rigid} refers to post-rigid registration, and \textit{Raw} indicates no motion correction. Metrics are shown as mean (standard deviation) except for NE\% and -J\% are shown as median (25\%, 75\%).}
    \resizebox{\columnwidth}{!}{
    \begin{tabular}{l@{\hskip 0.05in}c@{\hskip 0.05in}c@{\hskip 0.05in}c@{\hskip 0.15in}c@{\hskip 0.15in}c}
        \hline
        \textbf{Models} & \textbf{HA R$^2$ $\uparrow$} & \textbf{HA RMSE$\downarrow$} & \textbf{Residual$\downarrow$}  & \textbf{NE\%$\downarrow$} & \textbf{-J\%$\downarrow$}\\ \hline
        \textbf{0.1} & 0.566 (0.048) & 8.6 (2.7) & 15.0 (12.2) & 0.00 (0.00, 0.04) & 8.783 (6.774, 10.49) \\
        \textbf{0.5} & 0.603 (0.045) & 8.7 (2.4) & 16.7 (12.8) & 0.00 (0.00, 0.09) & 2.776 (2.051, 3.938) \\ 
        \textbf{2} & 0.674 (0.058) & 9.2 (2.3) & 20.0 (14.5) & 0.00 (0.00, 0.28) & 0.166 (0.113, 0.287) \\
        \textbf{5} & 0.749 (0.060) & 8.8 (2.1) & 21.2 (15.3) & 0.10 (0.01, 0.53) & 0.000 (0.000, 0.001) \\
        \textbf{10} & 0.793 (0.058) & 8.5 (1.8) & 23.0 (16.1) & 0.16 (0.03, 0.80) & 0.000 (0.000, 0.000) \\ 
        \textbf{20} & 0.823 (0.061) & 8.2 (1.6) & 24.1 (17.1) & 0.20 (0.04, 1.09) & 0.000 (0.000, 0.000) \\ 
        \textbf{50} & 0.857 (0.060) & 7.7 (1.6) & 25.7 (16.8) & 0.29 (0.08, 1.31) & 0.000 (0.000, 0.000) \\ 
        \textbf{100} & 0.875 (0.056) & 7.4 (1.5) & 26.2 (16.9) & 0.32 (0.10, 1.42) & 0.000 (0.000, 0.000) \\ 
        \textbf{200} & 0.891 (0.054) & 7.1 (1.5) & 26.7 (17.1) & 0.38 (0.10, 1.56) & 0.000 (0.000, 0.000) \\ 
        \textbf{300} & 0.896 (0.054) & 7.0 (1.5) & 27.1 (17.3) & 0.39 (0.13, 1.56) & 0.000 (0.000, 0.000) \\ 
        \textbf{Rigid} & 0.904 (0.060) & 7.1 (1.9) & 31.0 (19.2) & 0.47 (0.19, 2.03) & 0.000 (0.000, 0.000) \\ 
        \textbf{Raw} & 0.884 (0.061) & 7.4 (2.0) & 33.8 (21.3) & 1.16 (0.34, 2.69) & 0.000 (0.000, 0.000) \\ \hline     
    \end{tabular}
    }
    \label{tab:metrics}
\end{table}

\begin{figure*}[t]
\centering
\includegraphics[width=\textwidth]{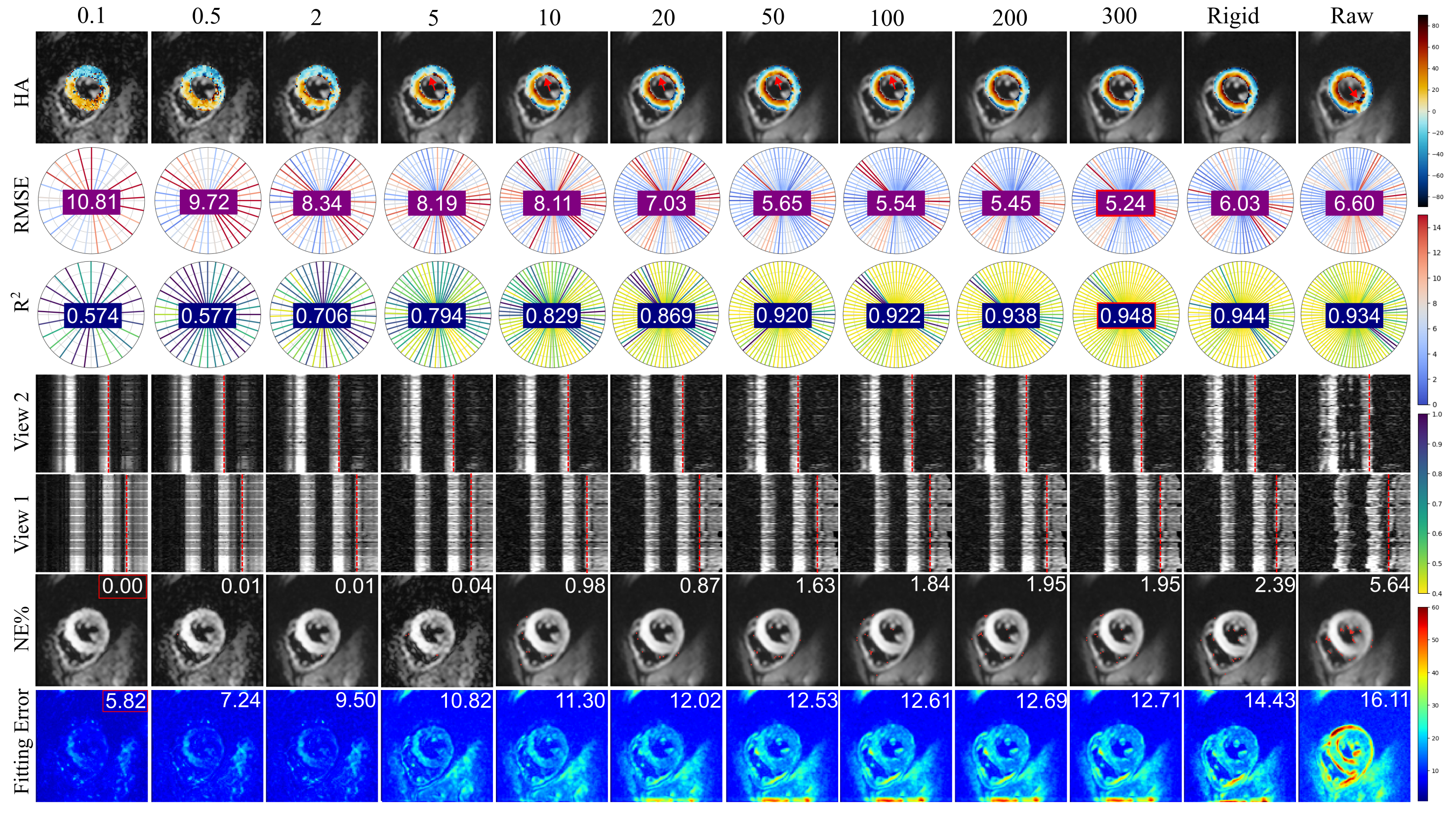}
\caption{Comparative results from a healthy volunteer are shown, with deformable registrations labelled by regularization weights. \textit{Rigid} and \textit{Raw} refer to rigid registration and no motion correction, respectively. HA metrics include root-mean-squared error (RMSE) and R$^2$ values for line profiles. \textit{View 1/2} represents horizontal and vertical perspectives of frames stacked along a third dimension. NE\%  of each method is labelled in the top-right corner. The mean fitting error is also displayed in the top-right corner. Colorbars on the right indicate HA, R$^2$ , RMSE, and fitting error, respectively.
} 
\label{Healthy}
\end{figure*}

In this study, we examined cDTI, which requires scanning a single slice multiple times under a free-breathing setting or among different breath-holds. To correct for the deformable motion inherent in these processes, both traditional \cite{nguyen2021free} and deep-learning based registration methods \cite{wang2024groupwise} have been proposed. We specifically evaluated the fitting error and concluded that this intuitive choice is a necessary but not sufficient metric for deformable motion correction. Relying solely on fitting errors can lead to deteriorated tensor quality and the loss of crucial physical features. We also explored the limitations and appropriate uses of fitting error in the context of cDTI, using a combination of deformable registration methods alongside clinical and physical metrics. We extended the findings to T1 mapping in the discussion, and underscored the need for caution when employing fitting error metric for motion correction in qMR.

\section{Methods}
\label{sec:format}
\subsection{Dataset}
We utilized in-house data from healthy volunteers as described in \cite{wang2024groupwise}, comprising 120 cases, each with over 70 frames of varying diffusion directions and b-values. Each frame was centrally cropped to 96$\times$96. We randomly selected 24 cases (20\%) for testing, 12 cases (10\%) for validation, and the remaining 84 cases (70\%) for training.

\begin{figure*}[t]
\centering
\includegraphics[width=\textwidth]{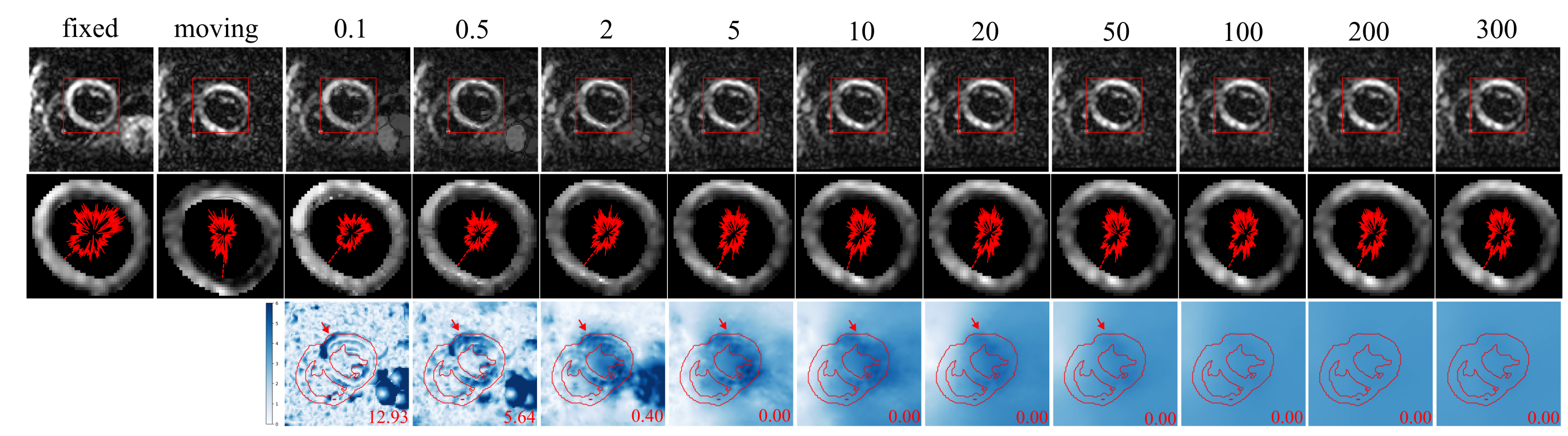}
\caption{Displacements of one moving frame. The title row shows deformable registration with different weight on the regularization with the fixed and moving frame. The second row visualizes the myocardium distribution, with mean intensity values plotted in polar coordinates. The third row illustrates the displacement magnitude of each frame, and includes the NE\% displayed in the right-bottom corner.} 
\label{HealthyDisp}
\end{figure*}

\subsection{Evaluation metrics}
Qualitatively, we stacked all frames along the third dimension to evaluate their alignment, displaying both horizontal and vertical views for comprehensive assessment. Besides the fitting error, we also used helix angle line profiles, percentage of the negative eigenvalues, and the Jacobian determinant as other evaluation metrics in Figure \ref{EvaluationMetric}.

\textbf{a. Fitting error} We employed pixel-wise nonlinear least-squares fitting, as described by Basser et al.~\cite{basser2011microstructural}, to estimate the parameters. Error measurements were calculated using the standard deviation per pixel, averaging over the series.
\textbf{b. Percentage of negative eigenvalues} For each pixel, a rank-2 diffusion tensor was computed, yielding three positive eigenvalues to quantify the magnitude of water molecule movement within the tissue \cite{ferreira2014vivo}. However, challenges such as noise, residual motion, and mis-alignment artifacts can result in negative eigenvalues. We assessed the percentage of negative eigenvalues (NE\%) in the centrally cropped portion of the images.
\textbf{c. Percentage of negative Jacobian determinant} The Jacobian Determinant, a second-order tensor field, is derived from the spatial derivatives of the displacement fields. A negative Jacobian determinant indicates the presence of a ``folded" point within the displacement. We quantified this by computing the pixels with negative Jacobian determinants over all the pixels of the displacement fields.
\textbf{d. Helix Angle Line Profile} In healthy individuals, the helix angle (HA) reflects the changes in myocardial fiber orientation, typically showing a linear progression from the endocardium to the epicardium. The average HA gradient was quantified based on the alignment degree relative to the myocardial wall thickness percentage. Radial line profiles with linear regression showing negative slopes and an R$^2$ value greater than 0.3 were included in the analysis \cite{gorodezky2018diffusion}. The R$^2$ and root-mean-squared-error (RMSE), as well as the steepness of these fittings were assessed. 

\subsection{Network Details}
We conducted the deformable registration by Transmorph \cite{chen2022transmorph} with different weights on the regularization to penalize the deformation. The brightest frame in each case was selected as the reference and all the other frames were registered to it in a pairwise manner. The frames were normalized to [0,1] by min-max method per frame as input and scaled back for tensor fitting. All the learning-based experiments were conducted on GTX 3090 using Ubuntu operating system. We fixed the batch size as 400, epoch number as 1000, learning rate as 10$^{-4}$ and the loss as the mutual information for all the experiments. We used the mutual information as the image similarity loss and Jacobian of the displacement field as the regularization term. We set the weight of the regularization over the mutual information as 0.1, 0.5, 2, 5, 10, 20, 50, 100, 150, 200, 300 respectively to check the performance of the fitting errors. The rigid registration was implemented using as low-rank based registration in Elastix \cite{wang2024low}. 

\section{Results}

\begin{figure}
\centering
\includegraphics[width=\columnwidth]{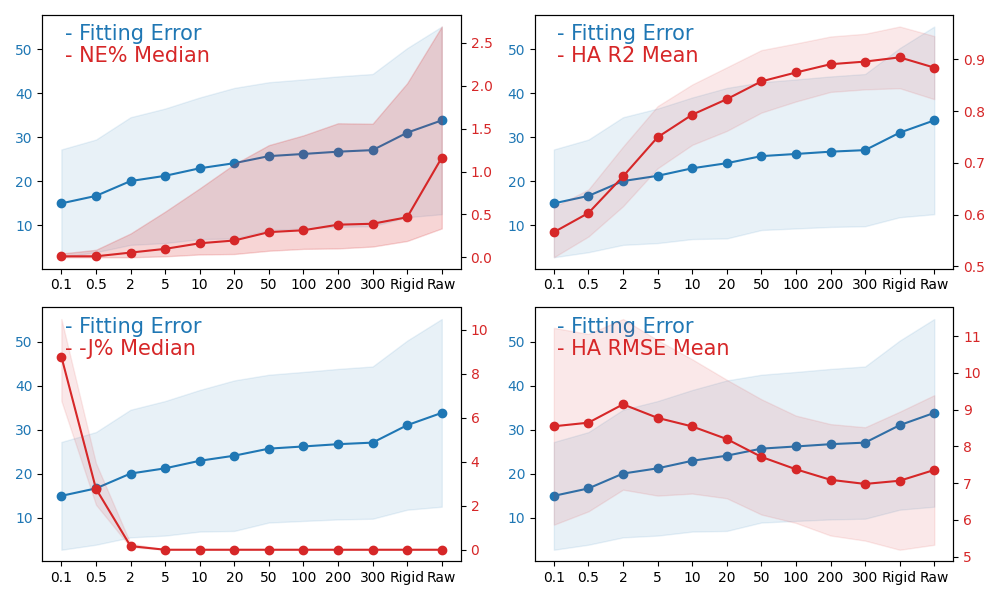}
\caption{The correlation between fitting error and other metrics.}
\label{Correlation}
\end{figure}

One healthy case is shown in Figure \ref{Healthy}. Small regularization weight shows perfect alignment, zero NE\% and less fitting error, but broken cardiomyocytes with low R$^2$ on HA fitting. As the regularization weight gradually increases, the linearity of HA increases with poorer alignment, higher NE\% and fitting error. Noted that the preferred rigid method in clinical settings, which avoids deformable registration that could distort the distribution, also reduces the fitting error and NE\%, while improving the HA line profiles. Therefore, while motion correction can reduce fitting errors, excessive deformable registration may compromise the quantitative features.

The rationale is illustrated in Figure \ref{HealthyDisp}. In pairwise registration, the moving frame exhibits respiratory shift relative to the fixed frame. With minimal regularization, excessive deformable registration causes the distributions and overall appearance in the moved frames to closely resemble the fixed frame. While this method allows the scales of different diffusion acquisitions to an exponential equation with lower errors, it compromises the accuracy.The HA analysis highlights that this reduction in fitting error is achieved at the expense of degrading the fitting tensor.

Table \ref{tab:metrics} and Figure \ref{Correlation} show the statistical results over the test dataset. As the regularization weight increases, the fitting error also increases due to a larger penalty on the displacement field. When the weight reached 10, the negative Jacobian determinant was eliminated. Improved R$^2$ and lower RMSE of HA line profiles indicate better alignment of cardiomyocytes as fitting errors increase. The NE\% shows a positive correlation with fitting error because excessive deformable registration causes the distributions to gather similarly, distorting the embedded tensor, yet results in moved images with fewer NE\%. The analysis found that fitting error positively correlates with the presence of NE\% and negatively correlates with the quality of helix angle line profiles. 

In cDTI, while motion correction can reduce fitting errors and enhance the HA line profiles, excessive registration may compromise the physiological features of the myocardial tissue, leading to deceptively improved statistical alignments.

\section{Conclusion and Discussion}

\begin{figure}
\centering
\includegraphics[width=\columnwidth]
{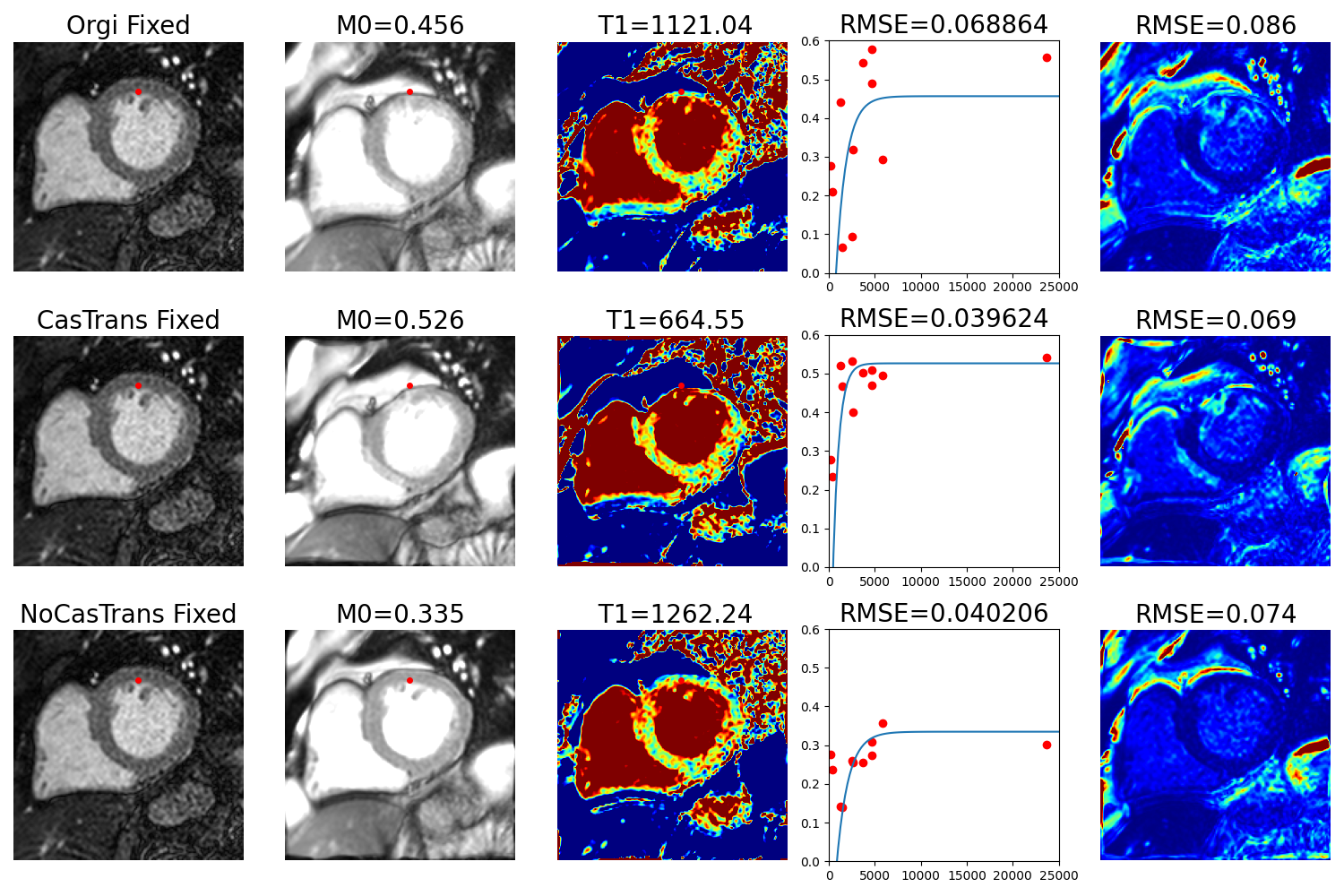}
\caption{
A case of free-breathing T1 mapping with the fixed frame was analyzed, focusing on the fitting parameters  M$_0$ and T1, along with a specific point in the myocardium, to assess both local and overall fitting errors. \textit{Raw}, \textit{CasTrans} and \textit{NoCasTrans} mean no motion correction, cascaded registration and the use of the first frame as the fixed in the pairwise registration. }
\label{T1}
\end{figure}

In this study, we have evaluated the reliability of the fitting error metric for deformable registration in cDTI. By integrating additional metrics such as clinical biomarkers like helix angle line profiles, and physical features such as negative eigenvalues and the Jacobian determinant of the displacement field, we caution against relying solely on fitting error as a definitive metric.

Under most settings where no mapping was acquired between the image pairs, further discussion on the promising metrics of deformable registration were conducted. A poor registration could improve the image similarity and the overlap on the deformed tissues, but show poor quality of moved anatomical images \cite{rohlfing2011image}. Different from anatomical acquisitions, qMR asks for a series of the images with different contrasts but the same anatomy embedded. Deformable registration is widely used in free-breathing \cite{nguyen2021free, arava2021deep,guyader2015influence,seif2016diffusion} or breath-hold scenarios \cite{wang2024groupwise,li2022motion,zhang2018cardiac, tao2018robust}. The delineation of tissue contours in such diverse and low signal-to-noise ratio (SNR) conditions is laborious and challenging.

As a down-stream process, it is intuitive to use the fitting error as a metric for qMR since the misalignment may increase the discrepancy between re-rendered images and the original images. 

The potential pitfalls of lower fitting errors masking poor image quality have been documented but not discussed \cite{arava2021deep}. We demonstrated this issue using a cascaded pairwise deformable registration network as shown in Figure \ref{T1}. This approach led to accumulated deformation errors, resulting in an implausible myocardium shape but paradoxically lower fitting errors at specific points and across the overall image. Such faulty registrations can alter the fitted parameters, misleadingly presenting reduced fitting errors.

Fitting error, defined as the standard deviations from the deformed points to the theoretical fitting lines, it a relative metric. Deformations that modify the overall fitting can reduce the fitting error but risk altering the fitting line itself, with higher precision but worse accuracy. Thus, while fitting error is a necessary component, it is not a sufficient metric for evaluating registration quality. We recommend using fitting error in conjunction with other metrics to comprehensively assess unsupervised deformable registration challenges. We aim to further validate this metric across various pathological conditions, additional centers, and different imaging modalities.

\section{Compliance with ethical standards}
This study was performed in line with the principles of the Declaration of Helsinki. Approval was granted by the Ethics Committee (National Research Ethics Service 10/H0701/112, approved October 2018). 

\section{Acknowledgments}
\label{sec:acknowledgments}

This study was supported in part by the ERC IMI (101005122), the H2020 (952172), the MRC (MC/PC/21013), the Royal Society (IEC\textbackslash NSFC\textbackslash 211235), the NVIDIA Academic Hardware Grant Program, the SABER project supported by Boehringer Ingelheim Ltd, NIHR Imperial Biomedical Research Centre (RDA01), Wellcome Leap Dynamic Resilience, UKRI guarantee funding for Horizon Europe MSCA Postdoctoral Fellowships (EP/Z002206/1), the British Heart Foundation grant (RG/F/23/110115, FS/19/22/34334) and the UKRI Future Leaders Fellowship (MR/V023799/1).
\bibliographystyle{IEEEbib}
\bibliography{reference}

\end{document}